\documentclass[
reprint,
onecolumn,
superscriptaddress,
amsmath,amssymb,
aps,
]{revtex4-2}

\usepackage{graphicx}
\usepackage{dcolumn}
\usepackage{bm}
\usepackage{hyperref}
\usepackage{url}
\usepackage{xcolor}
\usepackage[linesnumbered,ruled,vlined]{algorithm2e}
\usepackage{algpseudocode}
\usepackage{url}

\begin{document}


\title{Evolutionary vaccination dynamics under higher-order reinforcement pressure}

\newcommand{\YunnanAffil}{School of Statistics and Mathematics, Yunnan University of Finance and Economics, Kunming, Yunnan 650221, China}

\title{Evolutionary vaccination dynamics under higher-order reinforcement pressure}

\author{Yikang Lu}
\affiliation{\YunnanAffil}

\author{Ying Wang}
\affiliation{\YunnanAffil}

\author{Alfonso de Miguel-Arribas}
\affiliation{Zaragoza Logistics Center (ZLC), Zaragoza, 50018, Spain}

\author{Lei Shi}
\email{shi_lei65@hotmail.com}
\affiliation{\YunnanAffil}

\author{Yamir Moreno}
\affiliation{Institute for Biocomputation and Physics of Complex Systems (BIFI), University of Zaragoza, Zaragoza, 50018, Spain}
\affiliation{Department of Theoretical Physics, University of Zaragoza, Zaragoza, 50018, Spain}

\date{\today}

\begin{abstract}
Vaccination games in higher-order settings remain underexplored, despite their importance in shaping opinions and collective decisions. Here, we introduce a parsimonious behavioral-epidemiological model to evaluate how peer reinforcement pressure influences vaccination uptake. The framework consists of a two-layer multiplex: an epidemic layer governed by the SIR process on a square lattice, and a behavioral layer represented by a hypergraph of triadic interactions. Individuals update their vaccination strategy via imitation, modulated by a reinforcement parameter $\alpha$ when peer support is present. We find that higher-order structure alone induces clusters of vaccinated individuals that act as protective barriers. Low but nonzero reinforcement ($\alpha \approx 0.5$) maximizes coverage and suppresses outbreaks, while both negligible ($\alpha \approx 0$) and moderate ($\alpha > 0.1$) reinforcement reduce uptake, as excessive confirmation lowers adaptability and enables non-vaccinators to re-emerge. Our work bridges complex contagion theory with evolutionary game dynamics, offering insights into how contact structure and peer reinforcement jointly shape vaccination behavior.
\end{abstract}


\maketitle

\section{Introduction}
\label{sec:intro}

Higher-order interactions (HOIs) are increasingly recognized as essential components of human social behavior~\cite{mayfield2017higher, basu2018iterative, guo2021evolutionary}. Unlike simple pairwise exchanges, HOIs involve the simultaneous participation of three or more individuals, giving rise to collective effects that cannot be captured by dyadic models alone. Such interactions underpin a range of phenomena—from peer reinforcement in behavioral adoption to coordination in small-group decision-making—and are especially relevant in domains like opinion dynamics, cooperation, and behavioral contagion. This perspective becomes particularly important when individual decisions have collective consequences, such as in public health and epidemic control.

Thus, the spread of infectious diseases remains a major challenge for modern societies, with outbreak containment a central concern for both policymakers and researchers. Over the past few years, HOIs have emerged as a promising lens for understanding and managing these dynamics. Research in this area generally focuses on two key aspects: (i) how HOIs influence transmission~\cite{wang2024epidemic, ferraz2024contagion}, and (ii) how they affect the success of intervention strategies such as vaccination.

In studies of transmission mechanisms, various complex network structures have been explored, including clustered networks~\cite{ritchie2014higher}, community networks~\cite{ma2024impact}, multilayer networks, and higher-dimensional small-world networks~\cite{wang2022epidemic}. Iacopini et al.\cite{iacopini2019simplicial} introduced the simplicial contagion model (SCM), which captures both pairwise and group-level spreading by representing social interactions as simplicial complexes. Their work demonstrated that HOIs can lead to discontinuous phase transitions and bistability, where outbreak outcomes critically depend on the initial fraction of infected individuals. Matamalas et al.\cite{matamalas2020abrupt} extended this approach using a discrete-time SIS model on simplicial complexes, revealing that group-level (2-simplex) transmission fundamentally alters the system's dynamics through microscopic Markov chain approximations. Further advancing this perspective, Higham and de Kergorlay~\cite{higham2021epidemics} proposed a hypergraph-based SIS model in which contagion dynamics depend on group size and nonlinear infection probabilities. Their spectral analysis identified rigorous thresholds for epidemic persistence, providing insights into the difference between biologically driven and behaviorally mediated transmission. Gu et al.\cite{gu2024epidemic} studied HOIs in mobile populations and found that larger numbers of initial infection seeds trigger spatially localized interactions, which facilitate outbreaks. Arruda et al.\cite{ferraz2023multistability} introduced a binary-state threshold model on hypergraphs, demonstrating a variety of rich dynamical behaviors—including multi-stability, intermittency, and hybrid phase transitions—emerging from the interplay between HOIs and community structure. 

Yet understanding HOIs in transmission is only part of the picture: disease containment also depends on how individuals respond to preventive measures. In terms of prevention, vaccination remains the most effective strategy for epidemic control. However, widespread vaccine hesitancy—driven by concerns about safety, effectiveness, or cost—often undermines immunization efforts. A comprehensive understanding of epidemic control requires simultaneous consideration of both transmission processes and behavioral responses to interventions. Recent studies have applied HOIs frameworks to investigate how collective dynamics shape vaccination behavior.

Zou et al.\cite{zou2023information} examined vaccination decisions within family networks, comparing individual-based and family-based strategies. In the latter, a decision by one member leads to the vaccination of the entire family. Their results suggest that collective vaccination behaviors may be less effective than individual decision-making in suppressing disease spread. Jhun et al.\cite{jhun2021effective} proposed a hyperedge-targeted vaccination strategy within a simplicial SIS model, showing that immunizing high-risk hyperedges (based on H-eigenscores) effectively curbs outbreaks in uniform hypergraphs. Nie et al.\cite{nie2023voluntary} investigated how prioritizing group vs. individual payoffs influences vaccination uptake. Their findings indicate that under weak imitation dynamics, cost considerations dominate, while under strong social influence, group-oriented strategies promote higher coverage. Valdez et al.\cite{valdez2023epidemic} developed an SIRQ model on networks with cliques and showed that timely quarantine interventions can lead to both continuous and discontinuous suppression regimes, resulting in hybrid phase transitions and backward bifurcations.

Despite growing interest in complex contagion and higher-order interactions, such effects have received limited attention in vaccination game models—particularly at the behavioral decision-making stage. Most existing approaches still rely on dyadic imitation or incentive-based dynamics, with few exceptions such as the recent work by Nie et al.~\cite{nie2023voluntary}. 

In this study, we introduce a vaccination game on a spatially embedded hyper-lattice, where behavioral decisions are shaped by group-level (triadic) interactions and modulated by a reinforcement mechanism. Our framework couples this higher-order behavioral layer with a classical disease-spreading process occurring on a regular lattice. Within each hyper-edge, agents are assigned the roles of learner, reference individual, and observer. If the learner and observer hold different strategies, the learner may adopt the reference individual’s strategy according to a standard Fermi update rule. If the learner and observer share the same strategy, imitation is downscaled by a reinforcement factor governed by the parameter~$\alpha$. Through extensive numerical simulations, we identify an optimal range of~$\alpha$ that maximizes vaccination uptake for a given cost. Additionally, we observe the emergence of coherent spatial patterns—rectangular clusters of vaccinated individuals—under specific parameter regimes, revealing how higher-order feedback can structure collective behavior.

The remainder of this paper is organized as follows. In Section~\ref{sec:model}, we present the formulation of our vaccination game model. Section~\ref{sec:hmft} offers a Homogeneous Mean-Field Theory formulation of the model. Section~\ref{sec:results} provides a detailed analysis of the model’s behavior. Finally, Section~\ref{sec:discussion} summarizes the key findings and outlines directions for future research.

\section{Vaccination Game Model}
\label{sec:model}

As in any typical vaccination game, our model consists of an evolutionary process with two iterated stages: i) a decision-making stage in which individuals consider whether to get vaccinated under a game-theoretical framework, and ii) a disease transmission stage in which the epidemic spreads through the population. We refer to a \textit{season} as the full completion of these two consecutive stages. 

In the following, we describe the spatial structure where these dynamics unfold, along with the different elements of the vaccination game, i.e., the associated payoff structure, the decision-making rules, and eventually the contagion process model.

\subsection{Spatial structure}
\label{subsec:spatial}

We consider a population of $N$ agents, each occupying a node on a two-dimensional spatially embedded hyper-lattice\cite{lu2025enhancement,schawe2022higher} with periodic boundary conditions. As in classical lattice-based evolutionary games, each agent has four nearest neighbors. However, departing from the conventional pairwise interaction setting, each agent now participates in four distinct hyper-edges, each representing a group-level interaction of size three.

\begin{figure*}[!t]
	\centering
	\includegraphics[width=0.9\linewidth]{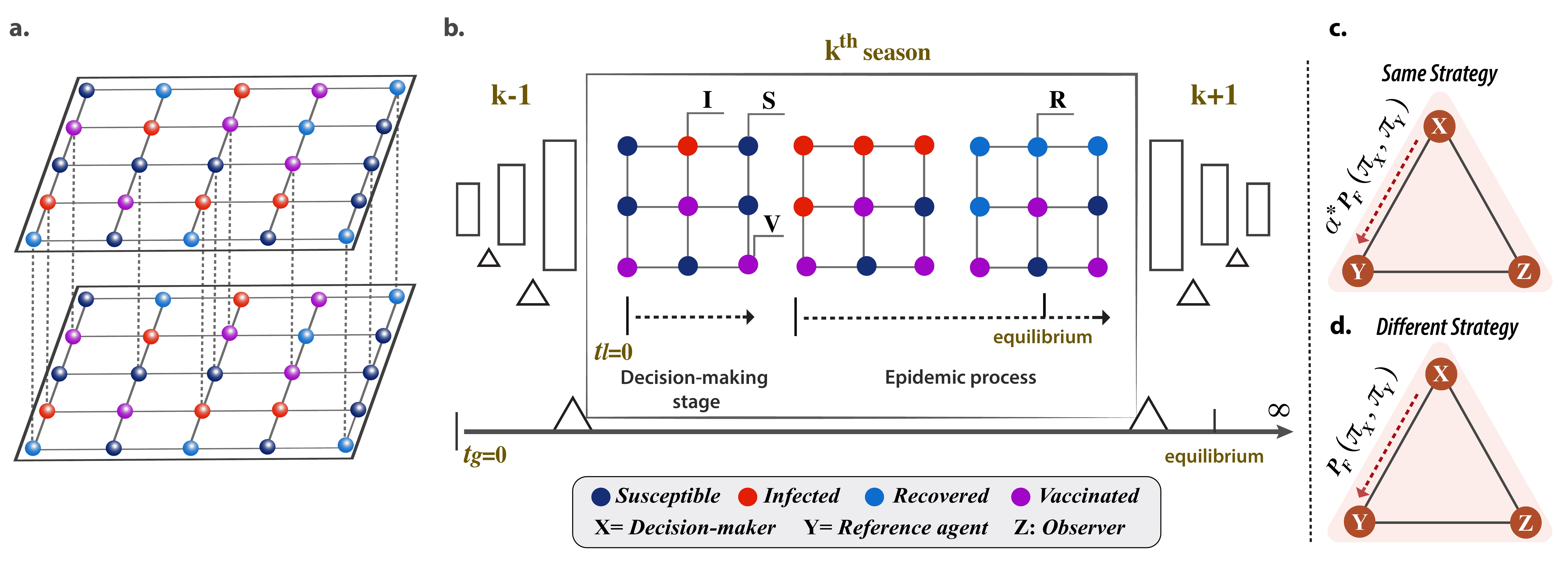}
	\caption{\textbf{Spatial structure and vaccination game scheme.} 
		\textbf{(a)} Schematic representation of the two-layer multiplex interaction structure. The top layer illustrates the behavioral network, modeled as a spatially embedded hyper-lattice. Each agent is embedded into four overlapping hyper-edges (triangles), each connecting the agent with two of its neighbors under periodic boundary conditions. The bottom layer depicts the disease transmission network, a regular square lattice through which infections propagate via pairwise contacts. 
		\textbf{(b)} Overview of the seasonal dynamics. Each season consists of two stages: a vaccination decision-making phase, where agents revise their strategy based on payoffs and peer influence within a hyper-edge; and an epidemic spreading phase, modeled by a Susceptible-Infected-Recovered (SIR) process restricted to non-vaccinated individuals. These stages repeat across seasons until the system reaches a behavioral steady state. 
		\textbf{(c)} Strategy updating when the observer and the reference agent share the same strategy, different from the focal agent's;  in this case, the imitation probability follows a standard Fermi rule. 
		\textbf{(d)} Strategy updating rule when the focal agent and the observer share the same strategy; here, the Fermi rule is rescaled by a factor $\alpha<1$, representing the reinforcement pressure parameter, reducing the probability of imitation.}
	\label{fig:scheme}
\end{figure*}

Each hyper-edge connects a focal agent with two of its adjacent neighbors, forming a triangular interaction group. Unlike a simple triangle of pairwise edges, the three agents within a hyper-edge participate in a simultaneous interaction. For any focal agent, the four hyper-edges are defined as follows (see Fig.~\ref{fig:scheme}):
\begin{itemize}
	\item North-East (NE) hyper-edge: includes the focal agent, their northern neighbor, and their eastern neighbor.
	\item South-East (SE) hyper-edge: includes the focal agent, their eastern neighbor, and their southern neighbor.
	\item South-West (SW) hyper-edge: includes the focal agent, their southern neighbor, and their western neighbor.
	\item North-West (NW) hyper-edge: includes the focal agent, their western neighbor, and their northern neighbor.
\end{itemize}
As a result, each hyper-edge has cardinality $3$, and each agent is a member of four overlapping hyper-edges that capture the local structure of higher-order interactions on the lattice.

\subsection{Strategies and payoffs}
\label{subsec:payoffs}

The decision-making process is framed within an evolutionary game-theoretic setting, which requires specifying the payoffs associated with each individual's possible strategy. In our model, the decision is binary: to vaccinate or not. We refer to this decision as a \textit{strategy}. These strategies are chosen at the beginning of each season, prior to the epidemic outbreak.

Vaccinated individuals incur a fixed cost $\pi_V=C_V$, reflecting the perceived cost of immunization (e.g., time, effort, side effects, or monetary expense). In contrast, individuals who opt out of vaccination bear no immediate cost during the decision-making stage. However, they are exposed to infection in the subsequent epidemic stage.

Non-vaccinated individuals who become infected during the disease transmission stage incur a cost $\pi_I=C_I$, representing the consequences of illness. Those who remain non-vaccinated but avoid infection altogether—commonly referred to as successful \textit{free riders}—incur no cost, $\pi_{SFR}=0$. For simplicity and without loss of generality, we normalize the infection cost to $C_I=1$, and define the relative vaccination cost as $c\equiv\frac{C_V}{C_I}$. This definition ensures normalized payoffs, $\pi\in[0,1]$.

The full payoff structure is thus summarized as follows:

\begin{equation}
	\label{eq:payoffs}
	\pi_X = \left \{
	\begin{array}{ll}
		0,       & \text{successful free rider (SFR),} \\
		c,     & \text{vaccinated individual } (\mathcal{V}), \\
		1,       & \text{infected individual } (\mathcal{I}).
	\end{array}
	\right.
\end{equation}

We emphasize that while individuals may end up in different epidemiological states (susceptible, infected, or recovered), their underlying strategy space remains binary: vaccinate or not. It is these strategies, not disease states, that are subject to imitation during the decision-making process.

\subsection{Decision-making process and updating rule}
\label{subsec:decision}

The decision-making process follows an evolutionary learning rule embedded in the hypergraph structure. At each iteration, a focal agent $l$, the \textit{learner}, selects uniformly at random one of the four hyper-edges in which they participate. This hyper-edge includes two other agents: one of them, denoted $r$, is randomly chosen as the \textit{reference agent}, whose strategy may be imitated by $l$; the other, denoted $o$, serves as an \textit{observer}, whose presence modulates the learner's likelihood of strategy updating.

This setup defines a higher-order learning mechanism based on triadic interactions. If the observer $o$ shares the same vaccination strategy $\sigma_r$, as the focal agent $l$, $\sigma_l$, the probability that $l$ adopts the strategy of the reference agent $r$ is reduced by a factor $\alpha<1$, called the reinforcement bias parameter, reflecting a lower propensity to change in the presence of like-minded peer support.  Conversely, if the observer $o$ holds a different strategy from $l$, the focal agent follows the standard Fermi rule for imitation.

The baseline Fermi probability function is defined as:
\begin{equation}
	P_F(\pi_l,\pi_r)=\frac{1}{1+\exp[-(\pi_r-\pi_l)/\kappa]},
	\label{eq:fermi}
\end{equation}
where $\kappa$ controls the level of noise in decision-making, with smaller values corresponding to more selective imitation. Following standard convention, we set $\kappa=0.1$ throughout.

The full updating rule, which defines the probability that agent $l$, holding strategy $\sigma_l$, adopts the strategy $\sigma_r$ of reference agent $r$, is given by:
\begin{equation}
	P(\sigma_i \rightarrow \sigma_j) = 
	\begin{cases}
		\alpha \cdot P_F(\pi_l,\pi_r), & \text{if } \sigma_l = \sigma_o, \\
		P_F(\pi_l,\pi_r),              & \text{if } \sigma_l \neq \sigma_o.
	\end{cases}
	\label{eq:update}
\end{equation}

Here, therefore, $\alpha\in[0,1]$ modulates the learning rate when reinforcement from a like-minded observer $o$ is present. This rule captures the influence of higher-order peer effects on vaccination decisions within localized social groups.

\subsection{Epidemiological model}
\label{subsec:epidemiological}

To simulate the spread of an influenza-like illness, we adopt a classical susceptible-infected-recovered (SIR) model with pairwise transmission. The epidemic unfolds immediately after each vaccination decision-making stage and only involves agents who chose not to vaccinate.

The SIR process consists of the following transitions:
\begin{align}
	\mathcal{S} + \mathcal{I} & \xrightarrow{\beta} \mathcal{I} + \mathcal{I}, \label{eq:infection}\\
	\mathcal{I} & \xrightarrow{\mu} \mathcal{R},
	\label{eq:recovery}
\end{align}
where $\beta$ is the transmission rate and $\mu$ is the recovery rate. Susceptible agents $\mathcal{S}$ can become infected $\mathcal{I}$ through pairwise contact with infected neighbors, while infected agents recover $\mathcal{R}$ at a constant rate $\mu$.

At each time step, the probability that a susceptible agent $i$ becomes infected is given by:
\begin{equation}
	P(\mathcal{S}\rightarrow\mathcal{I})=1-(1-\beta\Delta t)^{n_I},
\end{equation}
where $n_I$ is the number of infected neighbors of agent $i$.

The probability that an infected agent recovers is:
\begin{equation}
	P(\mathcal{I}\rightarrow\mathcal{R})=\mu\Delta t,
\end{equation}
with $\Delta t$ being a certain time step.

Vaccinated individuals are assumed to be fully protected and do not participate in the epidemic process. Therefore, only non-vaccinated susceptible agents are at risk of infection during the epidemic stage. The epidemic stage is iterated until no infected agents remain, after which the system returns to the decision-making phase for the next season.

\subsection{Simulations}
\label{subsec:simulations}

To investigate the dynamics of this model, we perform extensive Monte Carlo simulations of the coupled higher-order vaccination game and epidemic process.

The simulation procedure consists of four main steps:
\begin{enumerate}
	\item \textbf{Interaction structure:} A two-layer multiplex interaction network is constructed, consisting of a behavioral and a physical interaction layer. The behavioral layer is a spatially embedded hyper-lattice capturing higher-order interactions, while the physical layer is a regular square lattice representing pairwise disease transmission pathways. As this generative process is not stochastic, the contact structure is built once and reused across all simulations.
	\item \textbf{Behavioral stage:} Vaccination strategies are assigned randomly at season $s=0$. Each individual engages in the decision-making process by selecting one of their hyper-edges at random and updating their strategy according to the higher-order imitation rule (Eq.~\ref{eq:update}) based on their payoff (Eq.~\ref{eq:payoffs}). For successive seasons $s>0$, the vaccination outcome of season $s-1$ serves as the initial condition.
	\item \textbf{Epidemic stage:} The epidemic stage begins from the vaccination configuration $V(0)$ obtained in the previous stage. A small number $I_0$ of susceptible individuals are randomly selected to seed the epidemic. Disease transmission proceeds on the square lattice according to SIR dynamics (Eq.~\ref{eq:infection}, \ref{eq:recovery}), simulated using the continuous-time Gillespie algorithm~\cite{gillespie1977exact}. The SIR process terminates when all infected individuals transition to the recovered state (i.e., when $I(t\to\infty)=0$).
	\item \textbf{Seasonal iteration:} The behavioral and epidemic stages are iterated for $T$ seasons, until a steady-state has been reached in the decision-making evolutionary game dynamics. 
\end{enumerate}

To evaluate the long-term outcomes of the co-evolutionary process, we focus on two key observables: the vaccination coverage (VC), and the final epidemic size (FES). On the one hand, the vaccination coverage informs us about the fraction of the population adopting vaccination through the decision-making process, and is defined as:
\begin{equation}
	\langle v_{\text{st}}\rangle_{T_{eq}}=\frac{1}{N}\times\frac{1}{\Delta T}\sum_{s>T_{eq}}^TV_s.
\end{equation}
Here, $\langle \cdot\rangle_{T_{eq}}$ denotes the average over the ensemble of $\Delta T = T - T_{eq}$ seasons. We set $T_{eq}=2.5\times10^4$ and $T = 3\times10^4$, yielding $\Delta T = 5\times10^3$ seasons. $T_{eq}$ is defined as the season number from which measurements of the steady-state are performed and $V_s$ is the number of vaccinated individuals at season $s$. On the other hand, the final epidemic size measures the impact of the infectious disease on the population and is simply measured as:
\begin{equation}
	\langle r_{\text{st}}\rangle_{T_{eq}}=\frac{1}{N}\times\frac{1}{\Delta T}\sum_{s>T_{eq}}^TR_s(t\to\infty).
\end{equation}
Here, $R_s(t\to\infty)$ is the number of cumulative infected individuals that end up recovered by the end of the spreading ($t\to\infty$) at season $s$.

Additionally, given the stochastic nature of the simulated dynamics, all reported quantities $\langle v_{\text{st}}\rangle_{T_{eq}}$ and $\langle r_{\text{st}}\rangle_{T_{eq}}$ are averaged over $50$ independent realizations to ensure statistical robustness.

Unless otherwise specified, the rest of model parameters are set as follows: $\beta=0.46$, $\mu=0.333$, $N=10^4$, and $I(t=0)=10$. 

\section{Homogeneous Mean-Field Theory}
\label{sec:hmft}

To capture the average behavior of strategy adoption in the population, we develop a homogeneous mean-field approximation based on the triadic interactions taking place at the hyper-edges in the behavioral interaction layer. The core idea is that each individual's decision is influenced not just by pairwise comparisons but by higher-order social contexts, represented as hyper-edges (triplets of individuals). 

In general, for each triplet configuration (learner, reference individual, observer), the contribution to the dynamics of vaccination coverage is given by the probability of observing a certain triplet configuration times the probability of strategy update. Under a homogeneous mean-field assumption, the probability of selecting a specific hyper-edge configuration is approximated as a product of population fractions under the assumption of independence. 

We denote by $P(l \rightarrow r \mid o)$ the probability that a learner agent $l$ adopts the strategy of the reference individual $r$ in the presence of an observer $o$. This probability follows a Fermi-like update rule modulated by a reinforcement parameter $\alpha$ depending on the observer's strategy as defined in Eq. \ref{eq:update}. Based on all the different potential interactions, the update rules are defined as follows:

\begin{align}
	P(V \rightarrow F \mid V) &= \frac{\alpha}{1 + \exp\left[-\frac{-c}{\kappa}\right]}, \nonumber \\
	P(V \rightarrow I \mid V) &= \frac{\alpha}{1 + \exp\left[-\frac{-c + 1}{\kappa}\right]}, \nonumber \\
	P(V \rightarrow F \mid NV) &= \frac{1}{1 + \exp\left[-\frac{-c}{\kappa}\right]}, \nonumber \\
	P(V \rightarrow I \mid NV) &= \frac{1}{1 + \exp\left[-\frac{-c + 1}{\kappa}\right]}  \nonumber \\
	P(F \rightarrow V \mid V) &= \frac{1}{1 + \exp\left[-\frac{c}{\kappa}\right]}, \nonumber \\
	P(I \rightarrow V \mid V) &= \frac{1}{1 + \exp\left[-\frac{1 - c}{\kappa}\right]}, \nonumber \\
	P(F \rightarrow V \mid NV) &= \frac{\alpha}{1 + \exp\left[-\frac{c}{\kappa}\right]}, \nonumber \\
	P(I \rightarrow V \mid NV) &= \frac{\alpha}{1 + \exp\left[-\frac{1 - c}{\kappa}\right]}.
\end{align}

The full dynamics for the average vaccination coverage $v(t)$ is then obtained by summing over all possible update paths:
\begin{equation}
	\begin{aligned}
		\frac{dv}{dt} &= 
		- [v\;f\;v] \times P(V \rightarrow F \mid V) \\ &\quad
		- [v\;f\;(1\!-\!v)] \times P(V \rightarrow F \mid NV) \\
		&\quad 
		- [v\;i\;v] \times P(V \rightarrow I \mid V) \\ &\quad
		- [v\;i\;(1\!-\!v)] \times P(V \rightarrow I \mid NV) \\
		&\quad 
		+ [f\;v\;v] \times P(F \rightarrow V \mid V) \\ &\quad
		+ [f\;v\;(1\!-\!v)] \times P(F \rightarrow V \mid NV) \\
		&\quad 
		+ [i\;v\;v] \times P(I \rightarrow V \mid V) \\ &\quad
		+ [i\;v\;(1\!-\!v)] \times P(I \rightarrow V \mid NV).
	\end{aligned}
\end{equation}

Here, $[l\;r\;o]$ denotes the probability of randomly selecting a triad with types $l$ (learner), $r$ (reference), and $o$ (observer), e.g., $[v\;f\;v] = v \cdot f \cdot v$ under homogeneous mixing. Then $f=1-v-i$ is the fraction of successful free riders, $i$ is the fraction of infected individuals, and $v$ is the vaccination coverage.

To describe disease progression during the epidemic phase, we adopt the classical SIR model under homogeneous mixing. The time evolution of the susceptible, infected, and recovered fractions is given by:
\begin{equation}
	\begin{aligned}
		\frac{ds}{dt} &= -\beta s i, \\
		\frac{di}{dt} &= \beta s i - \mu i, \\
		\frac{dr}{dt} &= \mu i,
	\end{aligned}
\end{equation}
where, as previously defined, $\beta$ is the disease's transmission rate and $\mu$ is the recovery rate.

The total population is normalized such that $s(t)+v(t)+i(t)+r(t)=1$. The initial conditions are: $s(0)=1-v(0)-i(0)$, with  $r(0)=0$, $v(0)$ comes from the decision-making stage, and $i(0)=0.001$ as specified in Section \ref{subsec:simulations}.

The epidemic threshold is determined by the basic reproduction number $R_0$, which under a homogeneous or well-mixed population is given by $R_0=\beta/\mu$ \cite{weiss2013sir}. When $R_0>1$, an epidemic can spread, otherwise, the disease quickly dies out. The final epidemic size $r(\infty)$ satisfies the transcendental equation~\cite{fu2011imitation}:
\begin{equation}
	r(\infty)=(1-v(0))\left(1-e^{-R_0 r(\infty)}\right).
\end{equation}

Assuming vaccination provides perfect protection, the infection risk experienced by non-vaccinated individuals (i.e., the fraction who become infected) can be approximated by \cite{fu2011imitation}:

\begin{equation}
	w(v) = \frac{r(\infty)}{1 - v(0)} = 1 - e^{-R_0 r(\infty)}.
\end{equation}

This quantity represents the average probability that a free rider becomes infected, given a vaccination coverage $v(0)$. While it is a realized quantity after the epidemic has played out, it can be interpreted as an estimate of perceived infection risk and thus informs the social learning dynamics driving future vaccination behavior.

\begin{figure}[ht]
	\centering
	\includegraphics[width=0.9\linewidth]{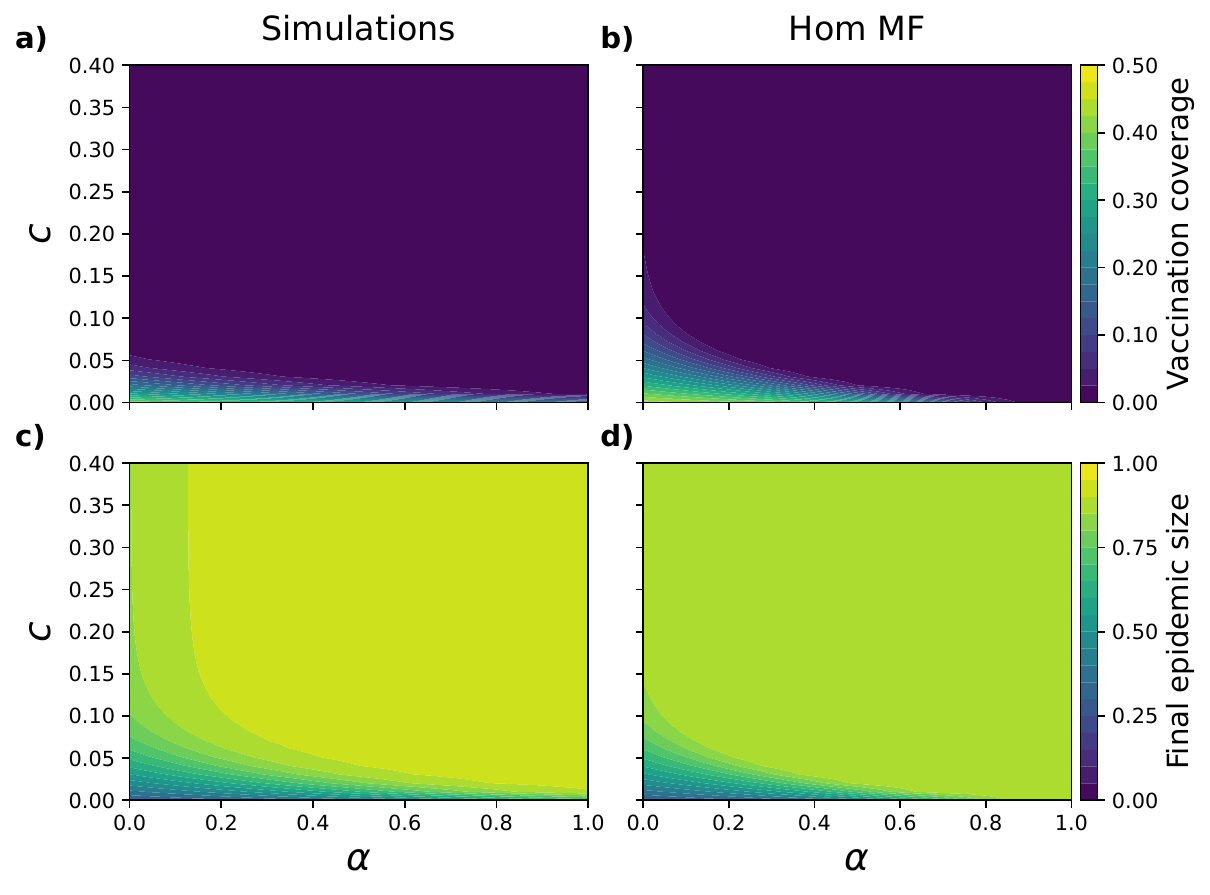}
	\caption{
		\textbf{Macroscopic behavior in a well-mixed setting.}
		Simulation results on a well-mixed system compared with analytical predictions from the homogeneous mean-field theory. Panels \textbf{a}) and \textbf{c}) show simulation outcomes on the well-mixed system, while panels \textbf{b}) and \textbf{d}) display the corresponding homogeneous mean-field theory predictions. Top panels depict vaccination coverage (VC) and bottom panels show the final epidemic size (FES).
	}
	\label{fig:hmft}
\end{figure}

\section{Results}
\label{sec:results}

We present the main results derived from the model exploration. First, we show the macroscopic phases of the system for VC and FES under varying control parameters $c$ and $\alpha$, both comparing simulations in a well-mixed setting against the homogeneous mean-field theory developed above, and next we display the macroscopic phases in the spatially-structured lattice multiplex. Afterwards, we delve into the system's microscopic behavior by identifying and analyzing characteristic configuration snapshots. Finally, we complement these results by exploring strategy transition probabilities under varied parameter settings.

\subsection{Macroscopic description}
\label{subsec:macro}

To gain a macroscopic understanding of how higher-order effects influence vaccination behavior and disease transmission, Figure~\ref{fig:hmft} presents simulation results on a well-mixed system alongside corresponding predictions from the homogeneous mean-field theory developed in Section \ref{sec:hmft}.

In Figure~\ref{fig:hmft}, the top panels (\textbf{a} and \textbf{b}) display vaccination coverage, while the bottom panels (\textbf{c} and \textbf{d}) show the final epidemic size. The left column presents simulation results on a well-mixed system, and the right column shows corresponding predictions from the homogeneous mean-field theory. Within well-mixed populations, we observe that for a fixed relative cost $c$, increasing the reinforcement parameter $\alpha$ reduces vaccination coverage, thereby increasing the final epidemic size. As expected, higher values of $c$ also lead to reduced vaccination uptake, in agreement with prior findings~\cite{fu2011imitation}.

\begin{figure*}[ht]
	\centering
	\includegraphics[width=1.0\linewidth]{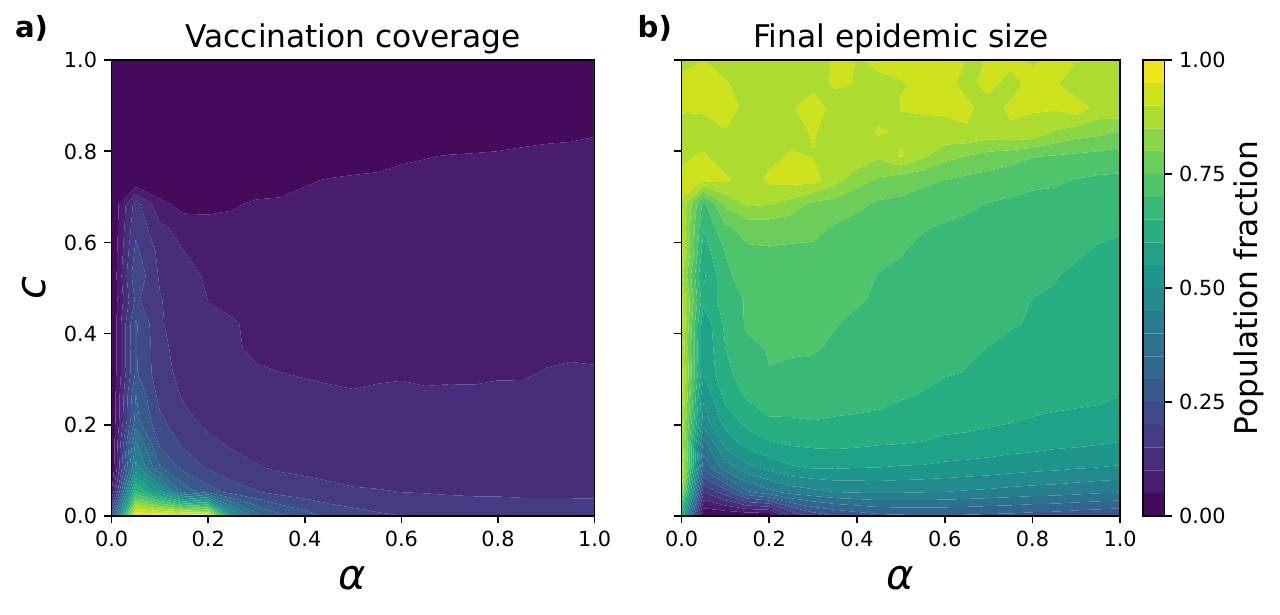}
	\caption{\textbf{Macroscopic behavior of the higher-order vaccination game.} Panel \textbf{a}: vaccination coverage, and Panel \textbf{b}: final epidemic size as functions of relative vaccination cost $c$ and reinforcement parameter $\alpha$ on a spatially embedded hyper-lattice. For fixed $c$, vaccination coverage peaks at an intermediate value of $\alpha$, indicating a non-monotonic relationship between higher-order influence and vaccine uptake in the structured lattice.}
	\label{fig:macro}
\end{figure*}

A direct comparison between simulation results and mean-field predictions reveals minimal discrepancies: absolute errors in vaccination coverage and final epidemic size are approximately $0.001\%$ and $0.0235\%$, respectively. These small deviations confirm that the mean-field approximation accurately captures the macroscopic behavior observed in simulations, as can be appreciated for the most part of the control parameter space.

The aforementioned results align well with the theoretical predictions under the assumption of homogeneous mixing. However, when network structure is incorporated, the outcomes exhibit notable deviations. Figure ~\ref{fig:macro} illustrates the vaccination coverage and final epidemic size in the lattice multiplex as functions of the relative vaccination cost $c$ and the reinforcement bias parameter $\alpha$, emphasizing the role of network topology on disease dynamics and individual decision-making. When the cost of vaccination is low, a higher proportion of individuals tend to adopt vaccination as a protective strategy, consistent with previous studies \cite{fu2011imitation}. As the reinforcement bias parameter $\alpha$ increases, vaccination coverage initially rises, peaking at $\alpha = 0.05$, and then gradually declines. At very low $\alpha$ values, individuals exhibit reluctance to adopt strategies from others when behaviors within a hyper-edge are inconsistent. As $\alpha$ increases, the influence of observed strategies weakens. When $\alpha$ exceeds approximately $0.3$, peer observation has a negligible effect, and social influence becomes marginal in shaping vaccination decisions. This non-monotonic response suggests that moderate levels of higher-order social influence can enhance vaccine uptake, whereas excessive reliance on self-confirmation may undermine collective immunity.

Our results reveal a marked contrast between well-mixed and structured populations. In homogeneous settings, low $\alpha$ values promote greater uptake but rarely suffice for disease elimination unless $c$ is low enough. Conversely, spatially structured hyper-lattices exhibit a region in the $(c,\alpha)$ space where herd immunity is achieved. Remarkably, this setting enables the emergence of an optimal value of $\alpha$ that maximizes vaccination coverage and fully suppresses the epidemic.

\subsection{Microscopic configurations}
\label{subsec:micro}

\begin{figure*}[ht]
	\centering
	\includegraphics[width=1.0\linewidth]{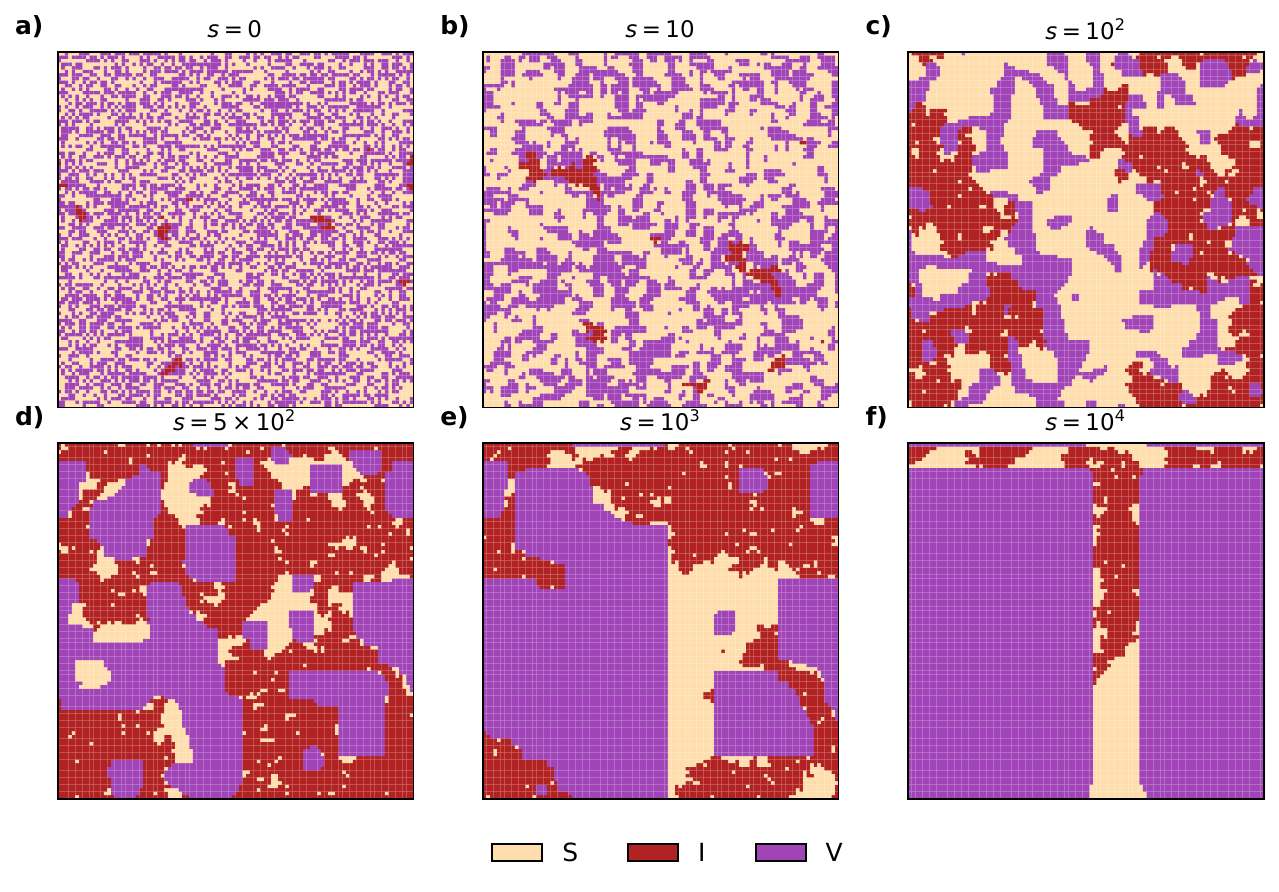}
	\caption{\textbf{Configuration evolutionary dynamics for $\alpha=0$.} Panels \textbf{a}) to \textbf{f}) depict, respectively, the microscopic evolution of the system for $\alpha=0$ and $c=0.05$ at seasons $s=0$, $10$, $10^2$, $5\times 10^2$, $10^3$, and $10^4$. At the steady-state, the system self-organizes into a giant rectangular cluster of vaccinated individuals.}
	\label{fig:micro_0}
\end{figure*}

After having a comprehensive picture of the macroscopic behavior of the system, we aim to gain a deeper understanding of vaccination attitudes by looking at the microscopic evolution of the system. To accomplish that, we depict a series of spatiotemporal configuration snapshots illustrating the dynamics of individual states.

Figure~\ref{fig:micro_0} presents evolutionary snapshots of the system's configuration under the conditions of relative vaccination cost $c=0.05$ and reinforcement bias parameter $\alpha=0$. At season $s=0$, vaccinated and susceptible strategies are randomly assigned to the individuals in the lattice. As the system evolves, both infected and vaccinated individuals gradually spread. Notably, vaccinated individuals self-organize into localized clusters that act as spatial barriers, impeding disease transmission. This emergent pattern reflects the collective behavioral dynamics in the absence of reinforcement bias resulting from the higher-order behavioral interaction.

\begin{figure}[ht]
	\centering	
	\includegraphics[width=1.0\linewidth]{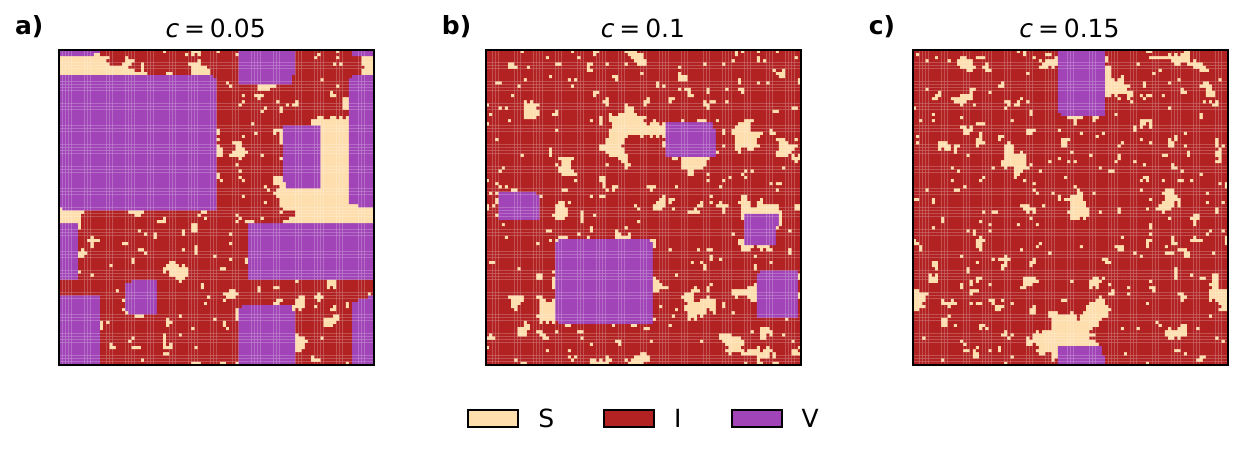}
	\caption{\textbf{Self-organized vaccination clusters at equilibrium.} Panels (\textbf{a}), (\textbf{b}), and (\textbf{c}) depict, respectively, the equilibrium configuration for the relative cost $c=0.05$, $0.1$, and $0.15$, with fixed $\alpha=0$ always (i.e., no reinforcement bias).
	}
	\label{fig:squares}
\end{figure}

Extensive simulations reveal that vaccination attitudes self-organize into rectangular-like spatial clusters within the parameter regime defined by $\alpha=0$ and relative cost $c\in[0,0.15]$, as illustrated in Fig.~\ref{fig:squares}. Individual vaccination uptake is effectively promoted only when the relative cost of vaccination remains low. However, when $\alpha=0$, within a hyper-edge, individuals update their strategies only when the other two nodes adopt a different strategy. This interaction rule facilitates the effective self-organization of individuals into distinctive rectangular-like structures.

To complement this microscopic evolutionary dynamics shown in Figure~\ref{fig:micro_0} for $\alpha=0$, we also track microscopic configurations under a positive reinforcement parameter $\alpha>0$ in the Appendix (see Figures ~\ref{fig:micro_005} and \ref{fig:micro_01}). Qualitatively, the dynamical evolution and steady-state are the same, being the main difference the shape of vaccination clusters in the transient state, and a higher vaccination coverage in the steady-state, thanks to the reinforcement mechanism. 

\begin{figure*}[ht]
	\centering
	\includegraphics[width=1.0\linewidth]{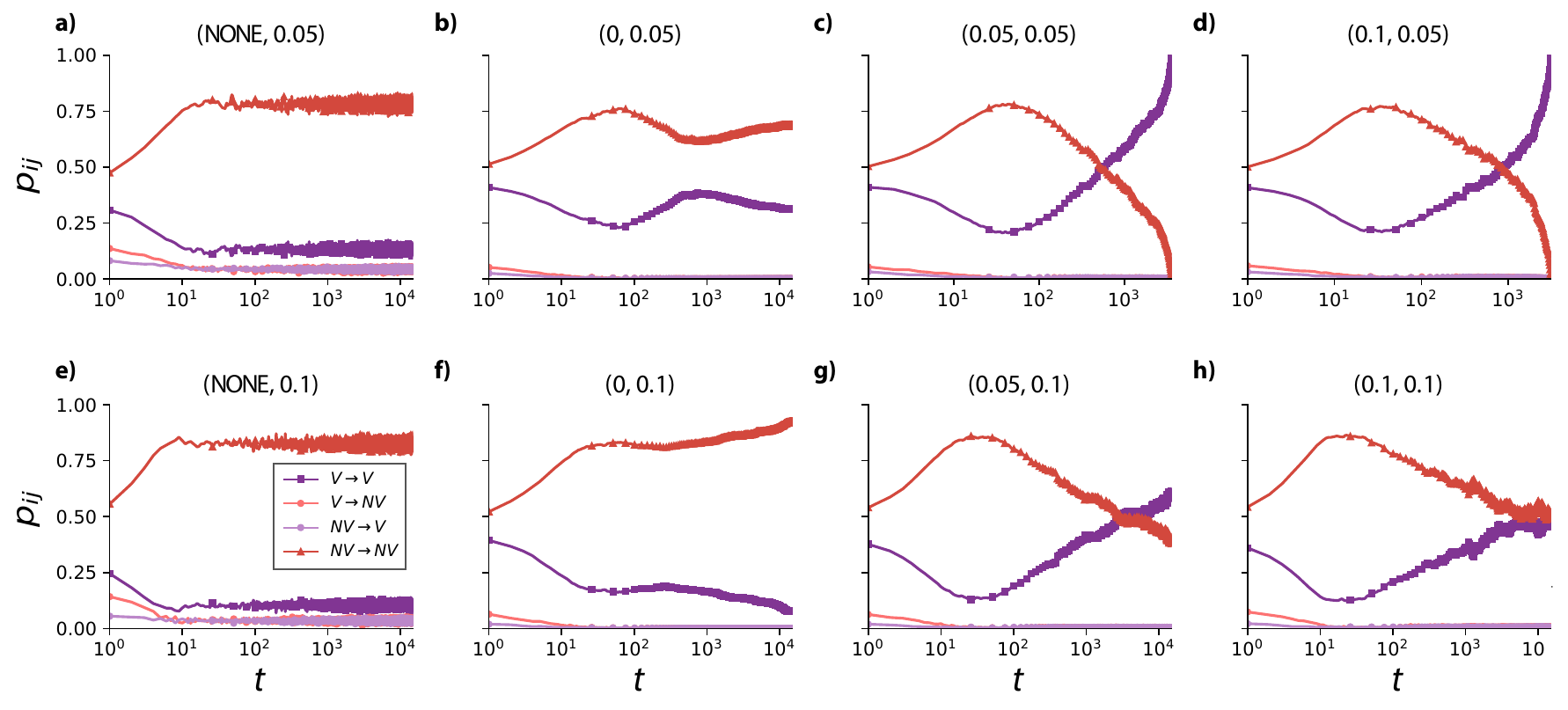}
	\caption{\textbf{Strategy switching with evolutionary time.} Strategy transition probabilities under various parameter combinations $(\alpha,c)$ are presented in panels (\textbf{a}) through (\textbf{h}), corresponding respectively to a baseline scenario with regular lattice structure and pairwise interactions, labeled as $(\mathrm{NONE}, 0.05)$, $(0, 0.05)$, $(0.05, 0.05)$, $(0.1, 0.05)$, $(\mathrm{NONE},0.1)$, $(0, 0.1)$, $(0.05,0.1)$, and $(0.1, 0.1)$. Panels (\textbf{a}) and (\textbf{e}) correspond to the baseline scenarios without spatial embedding or higher-order reinforcement ($\alpha=0$), i.e., the traditional results observed in regular square lattices. In the model, individuals adopt one of two strategies: vaccination ($V$) or refusal ($NV$), i.e., free-riding. We track behavioral dynamics across consecutive seasons by categorizing individuals based on their strategic transitions. Specifically, $V\to V$ denotes individuals who were vaccinated in the previous round and remained vaccinated; $V\to NV$ denotes those who switched from vaccination to refusal; $NV\to V$ denotes those who adopted vaccination after previously refusing; and $NV\to NV$ includes those who remained non-vaccinated. These four fractions fully characterize the evolution of individual behavior and allow us to quantify stability and change in strategic profiles.}
	\label{fig:transition}
\end{figure*}

\subsection{Seasonal strategy transition probabilities}
\label{subsec:transition}

To systematically explore how vaccination behavior responds to variations in cost and peer reinforcement influence, we analyze the evolution of strategic transitions across eight representative parameter settings, as shown in Fig.~\ref{fig:transition}. The top and bottom panels correspond to relative vaccination costs of $c=0.05$ and $c=0.1$, respectively. Moving from left to right, columns reflect increasing levels of higher-order/reinforcement pressure, with $\alpha=\text{NONE}$, $\alpha=0$, $\alpha=0.05$, and $\alpha=0.1$. The case $\alpha = \text{NONE}$ corresponds to the baseline scenario in which agents interact on a regular square lattice via pairwise interactions only, with no hyper-lattice structure and no reinforcement mechanism. In contrast, the case $\alpha = 0$ retains the hyper-lattice topology but disables reinforcement, thus isolating the structural effects of higher-order interactions from the reinforcement pressure modulation introduced by $\alpha$.

We define four transition probabilities between consecutive seasons:
\begin{itemize}
	\item $p_{V\to V}$: the fraction of vaccinated individuals who remain vaccinated from season $s$ to season $s+1$;
	\item $p_{V\to NV}$: the fraction of vaccinated individuals who switch to non-vaccination, thus free-riding;
	\item $p_{NV\to V}$: the fraction of non-vaccinated individuals who adopt vaccination;
	\item $p_{NV\to NV}$: the fraction of non-vaccinated individuals who remain non-vaccinated.
\end{itemize}
By construction, $p_{V\to V}+p_{V\to NV}+p_{NV\to V}+p_{NV\to NV}=1$, as each individual belongs to exactly one of these categories. These probabilities characterize the stability and flux of behavioral strategies in the population.

In the baseline scenario with $c=0.05$ and no higher-order interactions ($\alpha = \text{NONE}$), shown in Fig.~\ref{fig:transition}(a), the probability of remaining as non-vaccinated $p_{NV\to NV}$ quickly stabilizes at a high level, while the corresponding persistence for vaccinated individuals $p_{V\to V}$ somewhat below $0.2$. Transitions between strategies remain very limited: the probability $p_{NV\to V}$ of switching to vaccination hovers near $0.05$, indicating a weak incentive to adopt the cost-bearing strategy. This configuration reflects the self-reinforcing nature of non-vaccination and the structural stability of behavioral clusters.

When higher-order interactions are introduced through a finite reinforcement bias parameter ($\alpha\neq 0$), as in Fig.~\ref{fig:transition}(b), the transition landscape starts to change and showing some oscillatory behavior between the dominant transition probabilities. The vaccination persistence probability $p_{V\to V}$ initially declines slightly before recovering, while $p_{NV\to NV}$ is reduced compared to the baseline. Both switching probabilities $p_{NV\to V}$ and $p_{V\to NV}$ remain marginal, suggesting that higher-order effects suppress behavioral shifts and stabilize individual strategies.

As the strength of higher-order reinforcement increases further (Figs.~\ref{fig:transition}(c) and (d)), the vaccination persistence probability $p_{V\to V}$, after reaching a valley, steadily rises, ultimately reaching $1.0$ in a few thousand seasonal iterations. In this regime, all individuals eventually converge to and retain the vaccination strategy. Meanwhile, both $p_{NV\to NV}$ and the switching probabilities $p_{NV\to V}$ and $p_{V \to NV}$ decay to negligible levels, indicating a complete behavioral alignment and resolution of the social dilemma through higher-order reinforcement effects. 

When the cost is raised from $c=0.05$ to $c=0.1$, the baseline scenario given by pairwise interactions (Fig.~\ref{fig:transition}(e)) remain qualitatively similar to the lower-cost case, with a high $p_{V\to V}$, low $p_{NV\to V}$, and persistent non-vaccination clusters. 

Introducing the higher-order contact structure but without reinforcement bias $\alpha=0$ (Fig.~\ref{fig:transition}(f)) a similar wave-like behavior for the evolution of $p_{NV\to NV}$ and $p_{V\to V}$ trajectories is induced, but now, at later evolutionary steps, their gap is further increased, with persistent non-vaccinating almost dominating. 

Similarly as before, for the case of lower cost $c=0.05$, we find that this dominance is inverted when setting the finite and optimal reinforcement parameter value of  $\alpha=0.05$ (Fig.~\ref{fig:transition}(g)). Differently, though, this higher cost delays the arrival of the absorbing state of a fully vaccinated population.

However, and at variance with the case of $c=0.05$, now for $c=0.1$ and $\alpha=0.1$ (Fig.~\ref{fig:transition}(h)), the non-vaccination persistence $p_{NV\to NV}$ resurges and now they are on pair at the steady-state. This reversal indicates a non-monotonic relationship between reinforcement strength, as given by $\alpha$, and pro-vaccine behavior, and thus an optimal value of $\alpha$ beyond which, the efficacy of such interactions in sustaining pro-vaccination behavior is reduced.

In summary, higher-order interactions through the reinforcement bias parameter ($\alpha>0$) promote self-organizing behavioral stabilization and local consensus formation. For moderate $\alpha$, these effects support vaccination dominance by increasing persistence and reducing behavioral volatility. In particular, a near-complete dominance of vaccination is achieved when $\alpha$ surpasses $0.05$. Yet, the effect is non-monotonic: beyond a certain threshold, $\alpha=0.1$, stronger reinforcement can hinder global coordination by over-stabilizing fragmented behavioral basins. These findings underscore the complex, structure-sensitive role of peer reinforcement influence in collective health behaviors and suggest that optimal reinforcement levels may be necessary to overcome strategic inertia in public health interventions.

\section{Discussion}
\label{sec:discussion}

Human decision-making is shaped by a variety of social, cognitive, and contextual influences. In the context of vaccination behavior, individuals often rely not only on personal cost-benefit assessments but also on social cues and peer reinforcement. Understanding how such higher-order social influences interact with disease dynamics is essential for informing public health strategies. In this study, we have integrated higher-order behavioral updating rules into a framework that also captures disease spread through pairwise interactions, allowing us to jointly examine the coevolution of vaccination behavior and an influenza-like illness transmission.

Our model is built on a two-layer multiplex structure, with one layer representing behavioral interactions and the other encoding infectious disease transmission pathways. In the behavioral layer, individuals are embedded in a hypergraph structure (a hyper-lattice), where each triadic hyper-edge involves a \emph{learner}, a \emph{reference individual}, and an \emph{observer}. If the learner's strategy differs from that of the reference individual, the probability of imitation is given by a transition probability modulated by the reinforcement parameter $\alpha$, provided that the learner and observer share the same strategy. This setup introduces a triadic learning mechanism that captures the stabilizing influence of social reinforcement. When the learner and observer do not share the same strategy, imitation proceeds according to the standard pairwise rule based on payoff comparison. Disease transmission unfolds independently of the higher-order structure, taking place on the second layer of the multiplex—modeled as a regular square lattice—following the classical Susceptible-Infected-Recovered dynamics.

By analyzing this coupled system in both a hybrid homogeneous setting and a spatially explicit hypergraph, we have shown that higher-order interactions play a non-trivial role in shaping vaccination dynamics. Simulations on the homogeneous mean-field model agree well with pairwise approximation predictions, lending support to the theoretical consistency of our approach. In the hypergraph implementation, we find that for a given relative vaccination cost (e.g., $c<0.4$), there exists an optimal range of the reinforcement parameter $\alpha$ that maximizes the prevalence of vaccination. This range of values lies in the low end of $\alpha$, and it quickly narrows as the relative cost increases. Overall, either too low or moderate-to-high reinforcement pressure is detrimental to widespread vaccination coverage. This non-monotonic effect highlights the delicate balance set by reinforcement pressures.

Further inspection of spatial patterns reveals that even in the absence of reinforcement effects ($\alpha=0$), the mere presence of a higher-order structure can induce nontrivial organization: vaccinated individuals tend to self-organize into coherent rectangular clusters. These structures emerge from localized coordination facilitated by the triadic embedding, despite the lack of modulated transition probabilities. As $\alpha$ increases, triadic feedback becomes active, and the system transitions into a regime of widespread, stable vaccination—yet beyond a certain threshold, reinforcement becomes counterproductive, suppressing strategic flexibility and leading to the resurgence of non-vaccinated individuals. This phenomenon underscores the importance of tuning social reinforcement mechanisms to avoid behavioral rigidity.

Despite the insights provided by this study, several limitations should be acknowledged. First, our assumption of higher-order interactions in behavioral updating is theoretical and lacks direct empirical validation in the context of influenza vaccination. While the model offers a plausible mechanism for peer influence, real-world data on triadic or group-based learning dynamics remain scarce. Nevertheless, recent research on COVID-19 vaccine hesitancy suggests that vaccine uptake may follow complex contagion dynamics—where decisions are influenced by multiple peers—lending indirect support to the relevance of higher-order behavioral models \cite{de2025interplay}. Second, our analysis focuses on regular hypergraph structures and does not consider the effects of topological heterogeneity. In more realistic social systems—characterized by varying degrees of connectivity, clustering, or modularity—higher-order interactions may interact with network structure in more complex ways. Future work should extend this framework to heterogeneous or empirically informed hypergraphs, and consider dynamic or context-dependent values of $\alpha$ to reflect evolving social norms.

In sum, our results demonstrate that higher-order social learning dynamics can significantly alter the landscape of vaccination behavior and its interplay with epidemic processes. These findings highlight the potential of incorporating structured social influence into models of health behavior, offering a richer perspective on how cooperative strategies can emerge and persist in populations facing infectious threats.

\section*{Acknowledgments}
L.S. acknowledges support from the Major Program of the National Fund of Philosophy and Social Science of China (Grant Nos. 22\&ZD158 and 22VRCO49), the Key Project of the National Natural Science Foundation of China (Grant No. 11931015), the National Natural Science Foundation of China (Grant No. 11671348), and the Key R\&D Program of Yunnan Province (Grant No. 202403AC100010). 
Y.K. acknowledges support from the Yunnan Fundamental Research Projects (Grant Nos. 202501AU070193), the Scientific Research Fund of the Yunnan Provincial Department of Education (Grant No. 2025J0579) and Yunnan University of Finance and Economics (Grant No. 2025D60).
Y.M was partially supported by the Government of Arag\'on, Spain, and ``ERDF A way of making Europe'' through grant E36-23R (FENOL), and by Ministerio de Ciencia, Innovaci\'on y Universidades, Agencia Espa\~nola de Investigaci\'on 
(MICIU/AEI/ 10.13039/501100011033) Grant No. PID2023-149409NB-I00.

\appendix
\renewcommand{\thefigure}{\Alph{section}\arabic{figure}}

\section{Supplementary microscopic results}
\label{app:micro}

\setcounter{figure}{0}

\begin{figure}[ht]
	\centering
	\includegraphics[width=1.0\linewidth]{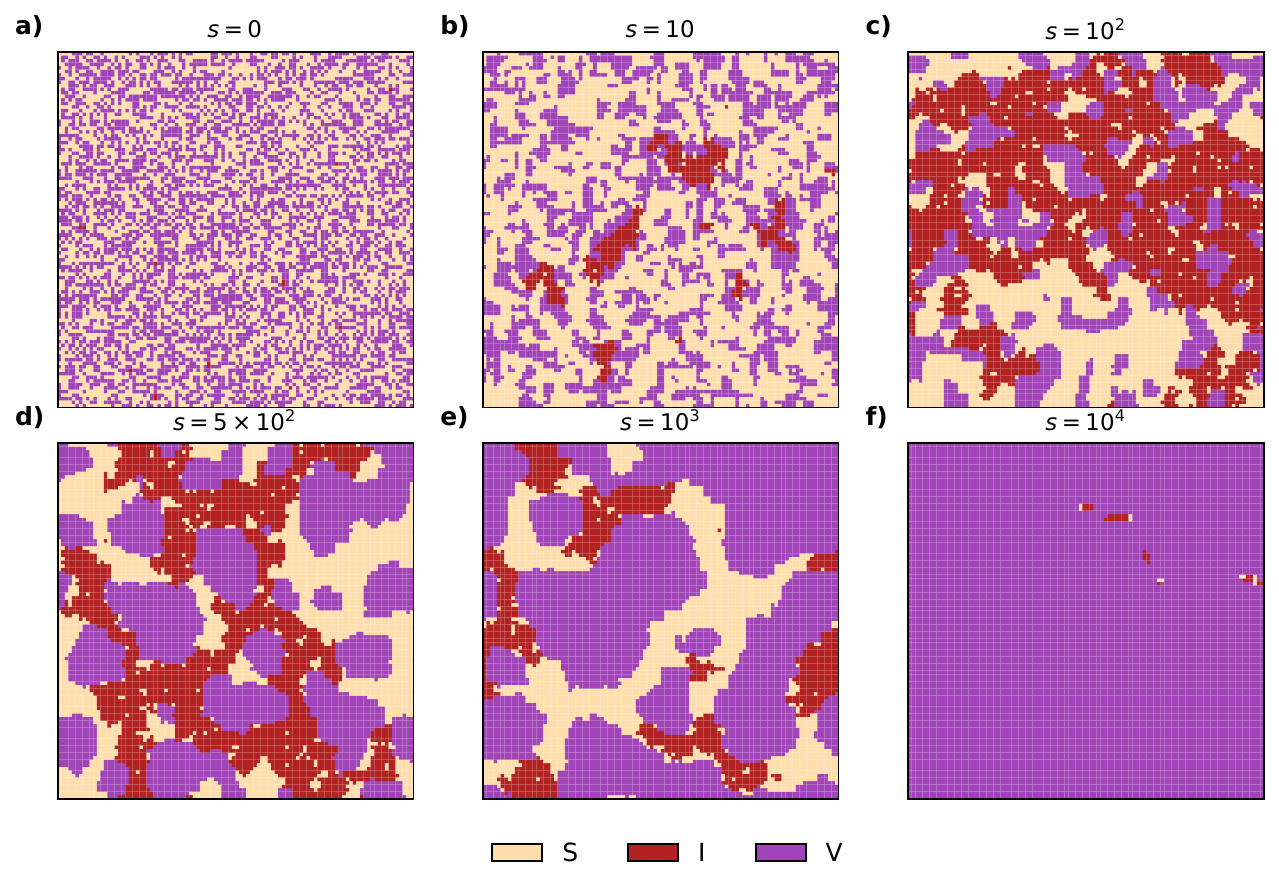}
	\caption{\textbf{Configuration evolutionary dynamics at $\alpha=0.05$.} Panels \textbf{a}) to (\textbf{f}), respectively, depict the microscopic evolution of the system for $\alpha=0.05$ and $c=0.05$ at seasons $s=0$, $10$, $10^2$, $5\times 10^2$, $10^3$, and $10^4$. Blue tiles represent susceptible individuals, while red and green represent, respectively, infected, and vaccinated individuals.}
	\label{fig:micro_005}
\end{figure}

\begin{figure}[ht]
	\centering
	\includegraphics[width=0.9\linewidth]{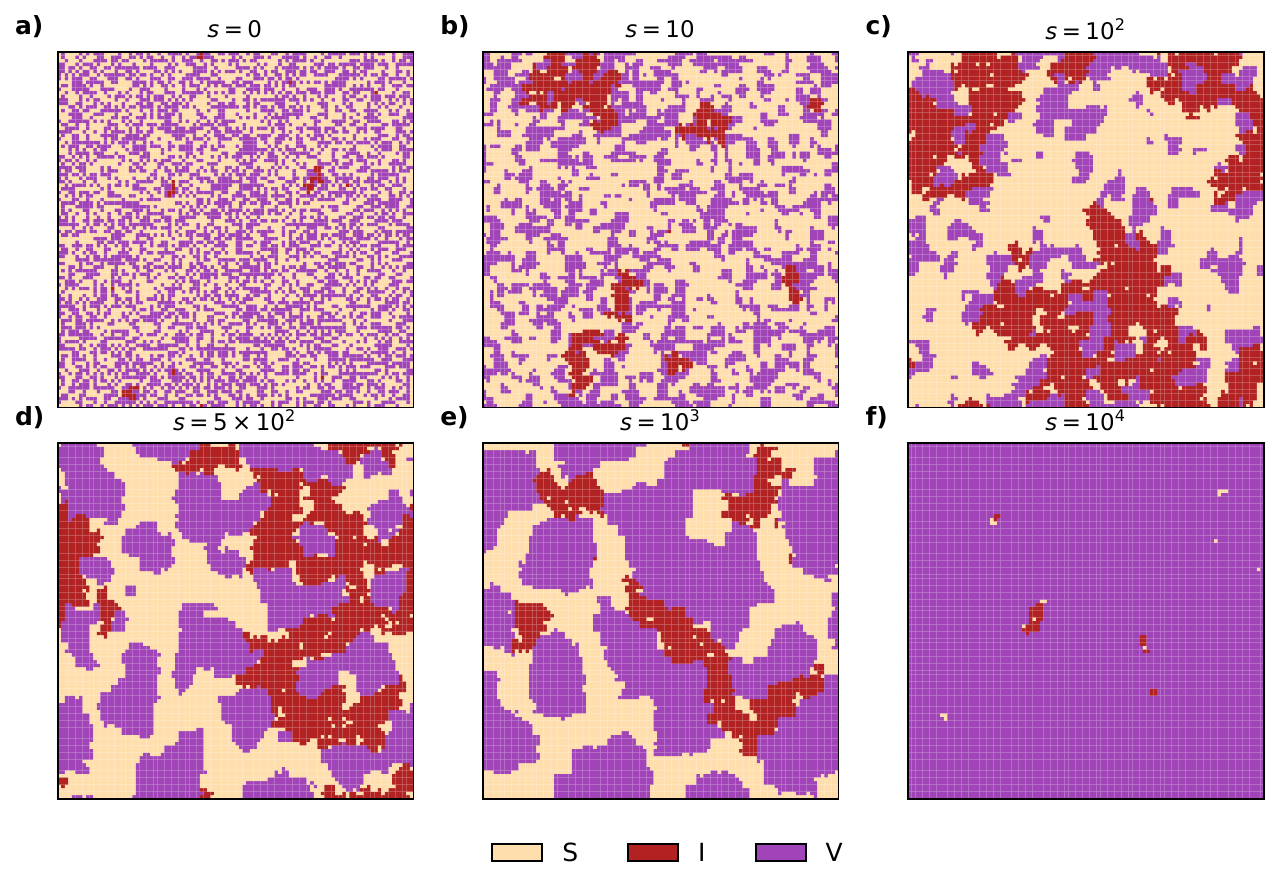}
	\caption{\textbf{Configuration evolutionary dynamics at $\alpha=0.1$.} Panels (\textbf{a})-(\textbf{f}), respectively, depict the microscopic evolution of the system for $\alpha=0.1$ and $c=0.05$ at seasons $s=0$, $10$, $10^2$, $5\times 10^2$, $10^3$, and $10^4$.}	
	\label{fig:micro_01}
\end{figure}

We offer supplementary results for the evolutionary dynamics of the system's microscopic configuration for $\alpha=0.05$ (see Fig.~\ref{fig:micro_005}) and for $\alpha=0.1$ (see Fig.~\ref{fig:micro_01}), thus complementing the configuration snapshots offered in the main text for $\alpha=0$ (Fig.~\ref{fig:micro_0}).

As $\alpha$ increases to $0.05$ from $\alpha=0$, the vaccination strategy becomes more prevalent. The initial distribution remains random, but over time, vaccinated individuals more rapidly form protective clusters, while the spread of infection is correspondingly limited to the susceptible gaps between these clusters. By season $s=500$, the system exhibits a more pronounced spatial segregation, with vaccinated regions effectively restricting the disease to narrow propagation paths.

When $\alpha$ is further increased to $0.1$, the adoption of vaccination is significantly enhanced, as shown on Fig.\ref{fig:micro_01}. At season $s=0$, infected individuals occupy only a small proportion of the population. By $t = 10$, the infection spreads more slowly compared to lower $\alpha$ scenarios, while vaccinated individuals begin to cluster. At season $s=100$, these clusters become larger, and the disease is primarily sustained among susceptible individuals located in the interstitial spaces between them. By $s=500$, the vaccinated population forms an extensive spatial barrier, substantially inhibiting disease transmission. This behavior resembles a negative feedback mechanism commonly observed in cooperative dynamics: as more individuals adopt vaccination, the opportunity for disease propagation is increasingly constrained. Eventually, vaccination becomes the dominant strategy across the population, effectively containing the epidemic and compressing transmission pathways to isolated susceptible channels.

\bibliographystyle{unsrt}
\bibliography{refs}

\end{document}